\documentclass[12pt]{iopart}

\usepackage{graphicx}

\begin{document}
\title[ ]{Nanomechanical Inverse Electromagnetically Induced Transparency and Confinement of Light in Normal modes}

\author{G. S. Agarwal$^{1}$ and Sumei Huang$^{2,3}$}

\address{$^1$Department of Physics, Oklahoma State University, Stillwater, Oklahoma 74078, USA}
\address{$^2$Department of Physics, University of California, Merced, California 95343, USA}
\address{$^3$Department of Electrical and Computer Engineering,
National University of Singapore, 4 Engineering Drive 3, Singapore 117583}
\eads{\mailto{girish.agarwal@okstate.edu}}
\date{\today}

\begin{abstract}
We demonstrate the existence of the phenomenon of the inverse electromagnetically induced
transparency (IEIT) in an opto mechanical system consisting of  a nanomechanical
mirror placed in an optical cavity. We show that two weak counter-propagating identical classical
probe fields can be completely absorbed by the system in the presence of a strong coupling field so
that the output probe fields are zero. The light is completely confined inside the cavity and the energy of the incoming probe fields is shared between the cavity field and creation of a coherent phonon and resides primarily in one of the polariton modes. The energy can be extracted by a perturbation of the external fields or by suddenly changing the $Q$ of the cavity.\\

\end{abstract}

\maketitle

\section {Introduction}

In recent years many analogs of the atomic interference effects~\cite{Harris,Fano}  have been extensively studied in optomechanical systems [OMS]. These include effects like electromagnetically induced transparency [EIT]~\cite{Agarwal1,Marquardt,Painter} and absorption~\cite{Qu,Kippenberg1}, Fano minima~\cite{Qu2013}. Such interference effects have led to the possibility of using OMS for storing optical pulses and more generally as optical memory elements~\cite{Wang1}. In EIT in OMS the cavity field vanishes and the output field is coherent and exactly equal to the input at EIT condition. In this paper we examine the possibility of confining light in the cavity with output equal to zero. We note that this can be considered as the inverse EIT (IEIT). The effect is similar to the coherent perfect absorption studied in the context of linear dielectrics~\cite{Wan}. In spite of the similarities there are differences as we deal with an active system and the entire effect is induced by the presence of a control field and hence it is appropriate to use the term IEIT in analogy to other electromagnetic field induced effects like EIT and EIA. The studies that we present in this paper deal with a system which is intrinsically nonlinear \cite{Review,Agarwalb}. Here the nonlinearity arises from the radiation pressure interaction between the mirror
and the cavity field. The dynamics of the cavity field and the mirror is completely described within the microscopic framework so that one is able to monitor dynamical changes in both the cavity field and the mechanical mirror. Another question of great importance is-where does the energy reside in the IEIT as it results from interference between coherent beams. Thus the question is-is the energy converted to heat or the energy resides in coherent vibrations, i.e., in the polaritons of the medium. Our microscopic approach shows that the coherent energy indeed gets converted to coherent internal vibrations of the mirror and the cavity field. We further find that at the IEIT conditions essentially only one of the two polariton modes is primarily excited. We present explicitly conditions on the dissipation of the mirror, the cavity damping, and
the power of the pump laser to obtain the frequencies at which the IEIT occurs. The frequency of the IEIT can
be tailored by changing the power and the frequency of the controlling electromagnetic field. This is where the nonlinearity of the radiation pressure interaction is useful, and thus our system can be used as a filter of frequencies which in turn would depend on the bandwidth of the incoming pump laser. We also note that since the IEIT arises from interference, the transmission of fields can be restored by changing the conditions for interference. The IEIT that we discuss is different from the EIA of references~\cite{Painter,Kippenberg1} which occurs when the control field is blue detuned and the EIA of~\cite{Qu} which can occur only for double cavity systems. Besides in IEIT the fields are completely confined inside the cavity.

The paper is organized as follows. In section 2, we describe the model under study, derive the quantum Langevin equations, and give the steady-state mean values. Then we linearize the quantum Langevin equations, make rotating wave approximation, and obtain the mechanical excitation and the cavity field at the probe frequency. In section 3, we calculate the output probe fields, show the IEIT in the output probe fields, derive the conditions for the IEIT occurrence, analyze where the energy goes and answer the question of the energy transfer to the normal modes when the IEIT occurs. Finally we present our conclusions in
section 4.

\section{Model}
\begin{figure}[htp]
\begin{center}
\scalebox{0.8}{\includegraphics{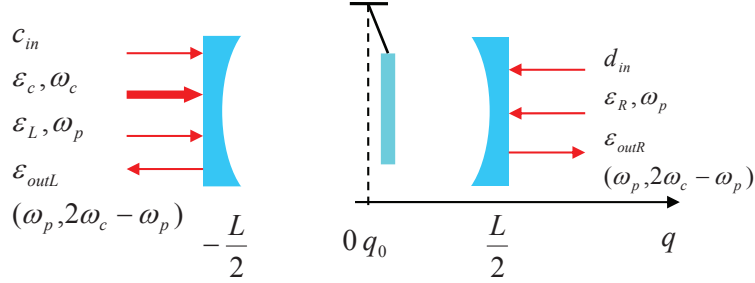}}
\caption{\label{Fig1} A double-ended cavity with a moving
nanomechancial mirror. The $q_{0}$ is the rest position of the moving mirror in the absence of radiation. These incoming fields $\varepsilon_{L}$ and $\varepsilon_{R}$ can interfere destructively under certain conditions, so that $\varepsilon_{outL}(\omega_{p})=\varepsilon_{outR}(\omega_{p})=0$.}
\end{center}
\end{figure}

 \noindent The studied system consists of a partially transmitting movable mirror located around the middle position of the Fabry-Perot cavity formed by two fixed mirrors with finite equal transmission~\cite{Jack}, as shown in Fig.~\ref{Fig1}. Such a system has been recently used \cite{Vitali} to demonstrate EIT at room temperature. The cavity field is driven by a strong coupling field with frequency $\omega_{c}$ and amplitude $\varepsilon_{c}$ from the left-hand side of the cavity. Meanwhile, two weak classical probe fields with identical frequency $\omega_{p}$ are sent into the cavity from the opposite sides of the cavity, respectively, their complex amplitudes are denoted by $\varepsilon_{L}$ and $\varepsilon_{R}$, respectively. The movable mirror makes small oscillations under the action of the radiation pressure force exerted by the photons within the cavity. In turn, the mechanical displacement $q$ modifies the cavity resonance frequency, represented by $\omega_{0}(q)$. We assume that the movable mirror is placed at the node of the frequency $\omega_{0}(q)$ of the cavity field, thus $\omega_{0}(q)$ depends linearly on the mechanical displacement $q$, $\omega_{0}(q)=\omega_{0}+g_{0}q$, where $\omega_{0}$ is the resonance frequency of the cavity in the absence of the moving mirror, the optomechanical coupling rate $g_{0}$ depends on the transmission $\mathcal{T}$ of the movable mirror $g_{0}=\frac{\sin{(2kq_{0})}}{\sqrt{(1-\mathcal{T})^{-1}-\cos^2(2kq_{0})}}\left(-\frac{\omega_{0}}{L/2}\right)$ \cite{Meystre}, where $k$ is the wave vector of the cavity field, $q_{0}$ is the rest position of the moving mirror in the absence of radiation. Here, the macroscopic movable mirror is treated as a quantum oscillator with effective mass $m$ and resonance frequency $\omega_{m}$. The Hamiltonian of the system in the rotating frame at the frequency of the coupling field $\omega_{c}$ is
\begin{eqnarray}\label{1}
H&=&\hbar(\omega_0-\omega_c)c^{\dag}c+\hbar g_{0} c^{\dag}c q+\frac{1}{2}m\omega_{m}^2q^2+\frac{p^2}{2m}+i\hbar \varepsilon_{c}(c^{\dag}-c)\nonumber\\& &+i\hbar (\varepsilon_{L}c^{\dag}e^{-i\delta t}-\varepsilon_{L}^{*}ce^{i\delta t})+i\hbar (\varepsilon_{R}c^{\dag}e^{-i\delta t}-\varepsilon_{R}^{*}ce^{i\delta t}),\ \ \delta=\omega_{p}-\omega_{c},
\end{eqnarray}
where $c$ ($c^{\dag}$) is the bosonic annihilation (creation) operator of the cavity mode with commutation relation $[c, c^{\dag}]=1$, $p$ is the momentum operator of the movable mirror, the amplitude $\varepsilon_{c}$ of the coupling field is determined by the power $\wp$ of the coupling field $\varepsilon_{c}=\sqrt{\frac{2\kappa \wp}{\hbar\omega_{c}}}$, and $\delta$ is the detuning between the probe field and the coupling field. It will be more convenient to write the position and momentum operators of the mirror in terms of the annihilation operator ($b$) and creation operator ($b^{\dag}$) of the mirror with $[b, b^{\dag}]=1$, $q=\sqrt{\frac{\hbar}{2m\omega_{m}}}(b+b^{\dag})$, and $p=i\sqrt{\frac{\hbar m\omega_{m}}{2}}(b^{\dag}-b)$, the Hamiltonian of the system can be rewritten as
\begin{eqnarray}\label{2}
H&=&\hbar(\omega_0-\omega_c)c^{\dag}c+\hbar g c^{\dag}c(b+b^{\dag})+\hbar \omega_{m}(b^{\dag}b+\frac{1}{2})+i\hbar \varepsilon_{c}(c^{\dag}-c)\nonumber\\
& &+i\hbar (\varepsilon_{L}c^{\dag}e^{-i\delta t}-\varepsilon_{L}^{*}ce^{i\delta t})+i\hbar (\varepsilon_{R}c^{\dag}e^{-i\delta t}-\varepsilon_{R}^{*}ce^{i\delta t}),
\end{eqnarray}
where $g=g_{0}\sqrt{\frac{\hbar}{2m\omega_{m}}}$ is the frequency shift of the cavity field induced by the zero-point motion of the mirror. Using the Heisenberg equations of motion and taking into account the corresponding damping and noise terms, one gets the quantum Langevin equations for the operators of the mechanical and optical modes
\begin{eqnarray}\label{3}
\dot{b}&=&-igc^{\dag}c-i\omega_{m}b-\frac{\gamma_{m}}{2}b+\sqrt{\gamma_{m}}b_{in},\nonumber\\
\dot{c}&=&-i(\omega_{0}-\omega_{c})c-igc(b+b^{\dag})+\varepsilon_{c}+\varepsilon_{L}e^{-i\delta t}\nonumber\\& &+\varepsilon_{R}e^{-i\delta t}-2\kappa c+\sqrt{2\kappa}c_{in}+\sqrt{2\kappa}d_{in}.
\end{eqnarray}
Here $\gamma_{m}$ is the mechanical damping rate due to the coupling of the movable mirror to the thermal environment, $2\kappa$ is the cavity photon decay rate due to transmission losses through each end mirror of the cavity, $b_{in}$ is the thermal noise on the movable mirror with zero mean value, $c_{in}$ ($d_{in}$) is the input quantum vacuum noise operator coming from the left- (right-) hand side of the cavity with zero mean value.

In the absence of the probe fields $\varepsilon_{L}$ and $\varepsilon_{R}$, the mean values in the steady state can be obtained from (3) by factorization $\langle cb\rangle=\langle c\rangle\langle b\rangle$ etc,
\begin{equation}\label{4}
\langle b\rangle=b_{s}=-\frac{ig|c_{s}|^2}{\frac{\gamma_{m}}{2}+i\omega_{m}},\ \
\langle c\rangle=c_{s}=\frac{\varepsilon_{c}}{2\kappa+i\Delta},
\end{equation}
where $\Delta=\omega_{0}-\omega_{c}+g(b_{s}+b_{s}^{*})$ denotes the effective detuning between the cavity field and the coupling field, including the frequency shift caused by the mechanical motion, $b_{s}$ determines the mechanical displacement at the steady state, and $c_{s}$ is the cavity field amplitude at the steady state.

To solve the nonlinear coupled equations (\ref{3}), we write each operator as the sum of the mean value and the small quantum fluctuation around this mean value i.e., $b=b_{s}+\delta b$ and $c=c_{s}+\delta c$, where $\delta b<<|b_{s}|$ and $\delta c<<|c_{s}|$. Inserting them into Eq. (\ref{3}), keeping only the linear terms, we obtain the linearized quantum Langevin equations
\begin{eqnarray}\label{5}
\delta \dot{\tilde{b}}&=&-ig[c_{s}^{*}\delta \tilde{c}e^{-i(\Delta-\omega_{m})t}+c_{s}\delta \tilde{c}^{\dag}e^{i(\Delta+\omega_{m})t}]-\frac{\gamma_{m}}{2}\delta \tilde{b} +\sqrt{\gamma_{m}}\tilde{b}_{in},\nonumber\\
\delta \dot{\tilde{c}}&=&-2\kappa \delta \tilde{c}-igc_{s}[\delta \tilde{b}e^{-i(\omega_{m}-\Delta)t}+\delta \tilde{b}^{\dag}e^{i(\omega_{m}+\Delta)t}]\nonumber\\& & +\varepsilon_{L}e^{-i(\delta-\Delta)t}+\varepsilon_{R}e^{-i(\delta-\Delta)t}+\sqrt{2\kappa}\tilde{c}_{in}+\sqrt{2\kappa}\tilde{d}_{in},
\end{eqnarray}
where we introduced the slowly moving operators with tildes, $\delta b=\delta\tilde{b}e^{-i\omega_{m}t}$, $b_{in}=\tilde{b}_{in}e^{-i\omega_{m}t}$, $\delta c=\delta\tilde{c}e^{-i\Delta t}$, $ c_{in}=\tilde{c}_{in}e^{-i\Delta t}$, $d_{in}=\tilde{d}_{in}e^{-i\Delta t}$.
If the cavity is driven by a coupling field at the mechanical red sideband $\Delta=\omega_{m}$, the system is operating in the resolved sideband regime $\omega_{m}>>\kappa$, the mirror has a high mechanical quality
factor $\omega_{m}>>\gamma_{m}$, and the mechanical frequency $\omega_{m}$ is much larger than $g|c_{s}|$, the fast oscillating terms $e^{\pm 2i\omega_{m}t}$ in Eq. (\ref{5}) can be ignored, Eq. (\ref{5}) simplifies to
\begin{eqnarray}\label{6}
\delta \dot{\tilde{b}}&=&-igc_{s}^{*}\delta \tilde{c}-\frac{\gamma_{m}}{2}\delta \tilde{b}+\sqrt{\gamma_{m}}\tilde{b}_{in},\nonumber\\
\delta \dot{\tilde{c}}&=&-2\kappa \delta \tilde{c}-igc_{s}\delta \tilde{b}+\varepsilon_{L}e^{-ixt}+\varepsilon_{R}e^{-ixt} +\sqrt{2\kappa}\tilde{c}_{in}+\sqrt{2\kappa}\tilde{d}_{in},
\end{eqnarray}
where $x=\delta-\omega_{m}$, the detuning of the probe field from the cavity resonance frequency.
Then we examine the expectation values of the small fluctuations, and note that the mean values of the quantum and thermal noise terms are zero. We write the solution for mean values in the form
\begin{equation}\label{7}
\langle \delta\tilde{s}\rangle=\delta\tilde{s}_{+}e^{-ixt}+\delta\tilde{s}_{-}e^{ixt},
\end{equation}
where $s=b$ or $c$, $\delta\tilde{s}_{+}$ and $\delta\tilde{s}_{-}$ are the components of $\langle \delta\tilde{s}\rangle$ oscillating at $\omega_{p}-\omega_{c}$ and $\omega_{c}-\omega_{p}$ in the rotating frame at frequency $\omega_{c}$, respectively, then we obtain
\begin{equation}\label{8}
\delta \tilde{b}_{+}=-\frac{igc_{s}^{*}}{\frac{\gamma_{m}}{2}-ix}\delta\tilde{c}_{+},\ \
\delta \tilde{c}_{+}=\frac{\varepsilon_{L}+\varepsilon_{R}}{2\kappa-ix+\frac{G^2}{\frac{\gamma_{m}}{2}-ix}},
\end{equation}
where $G=g|c_{s}|$ is the effective optomechanical coupling rate, which is related to the power $\wp$ of the coupling field. We have treated Eq.~(\ref{3}) in the linearized approximation which so far has been quite common. Consequences of going beyond linear approximation have started being examined \cite{Marquardt, Gupta}. Drawing on the findings of the ref. \cite{Gupta} we expect that the nonlinearities can be used to tune the position where the IEIT occurs.

\section{The IEIT in the output fields}
\noindent In this section, we calculate the output fields at the probe frequency $\omega_{p}$ to bring out the IEIT phenomenon due to the interaction of the movable mirror with a strong coupling field and two weak probe fields.

The output fields of the two sides of the cavity can be derived by the input-output relations \cite{Milburn}. They yields
\begin{equation}\label{9}
\varepsilon_{out\alpha}+\varepsilon_{\alpha}e^{-ixt}=2\kappa \langle \delta \tilde{c}\rangle,\ \ \alpha=R,L.
\end{equation}
Similarly, we write the output fields as
\begin{equation}\label{10}
\varepsilon_{out\alpha}=\varepsilon_{out\alpha+}e^{-ixt}+\varepsilon_{out\alpha-}e^{ixt},\ \ \alpha=R,L,
\end{equation}
where $\varepsilon_{out\alpha+}$ is oscillating at frequency $\omega_{p}$ in the original frame, $\varepsilon_{out\alpha-}$ is oscillating at frequency $2\omega_{c}-\omega_{p}$ in the original frame. From Eqs. (\ref{9}) and (\ref{10}), we obtain the output fields at the probe frequency
\begin{equation}\label{11}
\varepsilon_{out\alpha+}=2\kappa\delta \tilde{c}_{+}-\varepsilon_{\alpha},\quad \alpha=R,L.
\end{equation}
We find if the conditions
\begin{eqnarray}\label{12}
\varepsilon_{R}&=&\varepsilon_{L},\nonumber\\
\gamma_{m}&=&4\kappa,\nonumber\\
x&=&\omega_{p}-\omega_{c}-\omega_{m}=\pm \sqrt{G^2-4\kappa^2},\ \ G\geq2\kappa,\nonumber\\
G^2&=&g^2\frac{2\kappa \wp}{\hbar \omega_{c}(4\kappa^2+\omega_{m}^2)},
\end{eqnarray}
are satisfied, then the output fields are zero
\begin{equation}\label{13}
\varepsilon_{outR+}=\varepsilon_{outL+}=0.
\end{equation}
Therefore, the input probe fields are fully absorbed by this nonlinear optomechanical system without being reflected or transmitted. This is the result of the destructive interference between the reflected light of say left going probe field and the transmitted light from the probe field on the right. Hence this optomechanical system can achieve the IEIT. To make $x=\pm \sqrt{G^2-4\kappa^2}$ real, the effective coupling rate should be not less than the total cavity photon decay rate, i.e., $G\geq2\kappa$. For $G=2\kappa$, the IEIT happens at $x=0$, which means that the probe fields are resonant with the optical cavity. Further, by adjusting the power of the coupling field, the optomechanical system can entirely absorb the incoming two probe fields with equal frequency at $\omega_{p}=\omega_{c}+\omega_{m}\pm \sqrt{G^2-4\kappa^2}$. Clearly the transmission can be restored by changing any of the three conditions in Eq. (\ref{12}). One simple way is by changing the relative phase between $\varepsilon_{L}$ and $\varepsilon_{R}$. The IEIT occurs under conditions which are very different from the conditions for EIT. This is expected as for confinement of light we need good absorption and thus for IEIT we need a condition $\gamma_{m}=4\kappa$ which is different from the one, $\gamma_{m}\ll\kappa$ for EIT.

Next we examine the question where the energy resides when the IEIT occurs. For this purpose, we examine the intracavity probe photon number $|\delta\tilde{c}_{+}|^2$ and the quantum excitation $|\delta\tilde{b}_{+}|^2$ in the movable mirror when the IEIT happens. Substituting $\varepsilon_{R}=\varepsilon_{L}$, $\gamma_{m}=4\kappa$, and $x=\pm \sqrt{G^2-4\kappa^2}$ into Eq. (\ref{8}), we find that the normalized intracavity probe photon number, defined as the ratio of the intracavity probe photon number $|\delta\tilde{c}_{+}|^2$ versus the intracavity probe photon number $|\frac{\varepsilon_{L}}{2\kappa}|^2+|\frac{\varepsilon_{R}}{2\kappa}|^2$ without the coupling field, is
\begin{equation}\label{14}
\frac{4\kappa^2}{|\varepsilon_{L}|^2+|\varepsilon_{R}|^2}|\delta\tilde{c}_{+}|^2=\frac{1}{2}.
\end{equation}
From Eq. (\ref{8}), when $\varepsilon_{R}=\varepsilon_{L}$, $\gamma_{m}=4\kappa$, and $x=\pm \sqrt{G^2-4\kappa^2}$, the normalized mechanical excitation is
\begin{equation}\label{15}
\frac{4\kappa^2}{|\varepsilon_{L}|^2+|\varepsilon_{R}|^2}|\delta\tilde{b}_{+}|^2=\frac{1}{2}.
\end{equation}
Therefore, when the IEIT occurs, the intracavity probe photon number is equal to the mechanical excitation. Thus the energy of the input probe fields are totally transferred into the cavity field and the movable mirror, which have equal excitation. Further both the cavity mode and the phonon mode are coherently excited.

\begin{figure}[htp]
\begin{center}
\scalebox{0.7}{\includegraphics{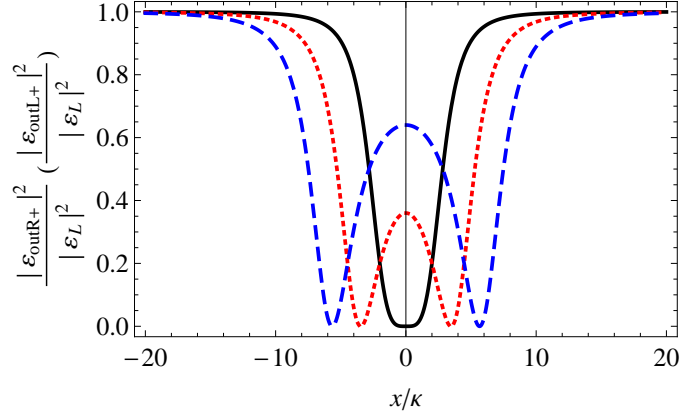}}
\caption{\label{Fig2} The normalized output probe photon number $|\frac{\varepsilon_{outR+}}{\varepsilon_{L}}|^2$ $(|\frac{\varepsilon_{outL+}}{\varepsilon_{L}}|^2)$ as a function of the probe detuning $x=\omega_{p}-\omega_{c}-\omega_{m}$. The solid, dotted, and dashed curves are for different pumping rates $G=2\kappa, 4\kappa, 6\kappa$.}
\end{center}
\end{figure}

\begin{figure}[htp]
\begin{center}
\scalebox{0.7}{\includegraphics{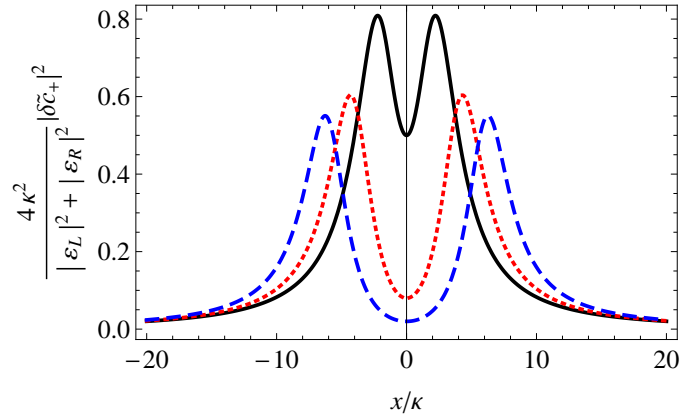}}
\caption{\label{Fig3} The normalized probe photon number $\frac{4\kappa^2}{|\varepsilon_{L}|^2+|\varepsilon_{R}|^2}|\delta\tilde{c}_{+}|^2$ in the cavity as a function of the probe detuning $x$. The solid, dotted, and dashed curves are for different pumping rates $G=2\kappa, 4\kappa, 6\kappa$.}
\end{center}
\end{figure}

\begin{figure}[htp]
\begin{center}
\scalebox{0.7}{\includegraphics{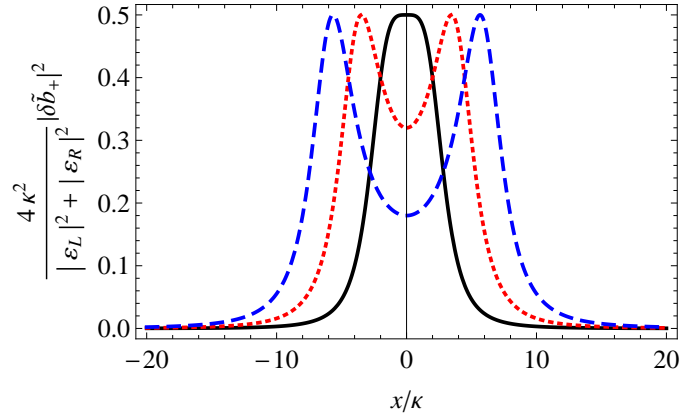}}
\caption{\label{Fig4} The normalized mechanical excitation $\frac{4\kappa^2}{|\varepsilon_{L}|^2+|\varepsilon_{R}|^2}|\delta\tilde{b}_{+}|^2$ as a function of the probe detuning $x$. The solid, dotted, and dashed curves are for different pumping rates $G=2\kappa, 4\kappa, 6\kappa$.}
\end{center}
\end{figure}

In the following, we numerically evaluate the output probe photon number, the intracavity probe photon number, and the mechanical excitation to show the effect of the power of the coupling field on the IEIT.
The normalized output probe photon number $|\frac{\varepsilon_{outR+}}{\varepsilon_{L}}|^2$ ($|\frac{\varepsilon_{outL+}}{\varepsilon_{L}}|^2$), the normalized probe photon number $\frac{4\kappa^2}{|\varepsilon_{L}|^2+|\varepsilon_{R}|^2}|\delta\tilde{c}_{+}|^2$ in the cavity, and the normalized mechanical excitation $\frac{4\kappa^2}{|\varepsilon_{L}|^2+|\varepsilon_{R}|^2}|\delta\tilde{b}_{+}|^2$ as a function of the probe detuning $x$ for different effective coupling rates ($G=2\kappa, 4\kappa, 6\kappa$) are shown in Figs.~\ref{Fig2}-\ref{Fig4}. For $G=2\kappa$, it is seen that when $x=0$, $\varepsilon_{outR+}=\varepsilon_{outL+}=0$, $\frac{4\kappa^2}{|\varepsilon_{L}|^2+|\varepsilon_{R}|^2}|\delta\tilde{c}_{+}|^2=\frac{1}{2}$, and $\frac{4\kappa^2}{|\varepsilon_{L}|^2+|\varepsilon_{R}|^2}|\delta\tilde{b}_{+}|^2=\frac{1}{2}$. For $G=4\kappa$ or $6\kappa$, one can see that when $x=\pm \sqrt{G^2-4\kappa^2}$, $\varepsilon_{outR+}=\varepsilon_{outL+}=0$, $\frac{4\kappa^2}{|\varepsilon_{L}|^2+|\varepsilon_{R}|^2}|\delta\tilde{c}_{+}|^2=\frac{1}{2}$, and $\frac{4\kappa^2}{|\varepsilon_{L}|^2+|\varepsilon_{R}|^2}|\delta\tilde{b}_{+}|^2=\frac{1}{2}$. These results are consistent with the analytical results (\ref{14}) and (\ref{15}). In addition, in the strong coupling regime $G>2\kappa$, two dips or two peaks appear in the output probe field, the intracavity probe field, and the mechanical excitation, this is the optomechanical normal mode splitting phenomenon~\cite{Kippenberg,Aspelmeyer} in the pump probe response of  the optomechanical system as is seen from Eq. (\ref{8}) which has poles at $x=\pm G-2i\kappa$. For $G=2\kappa$, the width and the location of the poles are comparable and this is the reason for broad structure at $x=0$ in the mechanical excitation. We next examine the question of energy transfer to the normal modes under the conditions of the IEIT. From Eqs. (\ref{6}) and (\ref{7}), we obtain the equations for the components $\delta\tilde{b}_{+}$ and $\delta\tilde{c}_{+}$
\begin{equation}\label{16}
-ix\left(\begin{array}{c}\delta\tilde{b}_{+}\\ \delta\tilde{c}_{+}\end{array}\right)=A\left(\begin{array}{c}\delta\tilde{b}_{+}\\ \delta\tilde{c}_{+}\end{array}\right)+\left(\begin{array}{c}0\\ \varepsilon_{L}+\varepsilon_{R}\end{array}\right),
\end{equation}
where $A$ is the $2\times2$ matrix
\begin{equation}\label{17}
A=\left(\begin{array}{cc}
-\frac{\gamma_{m}}{2} & -i g c_{s}^{*}\\
-igc_{s} & -2\kappa
\end{array}\right).
\end{equation}
\begin{figure}[htp]
\begin{center}
\scalebox{0.7}{\includegraphics{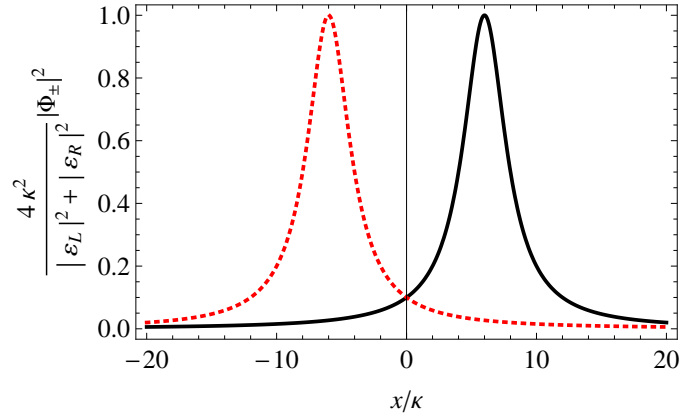}}
\caption{\label{Fig5} The normalized modulus square of the normal modes $\frac{4\kappa^2}{|\varepsilon_{L}|^2+|\varepsilon_{R}|^2}|\Phi_{\pm}|^2$ as a function of the probe detuning $x$ for the pumping rate $G=6\kappa$. The solid curve is for $\frac{4\kappa^2}{|\varepsilon_{L}|^2+|\varepsilon_{R}|^2}|\Phi_{+}|^2$, the dotted curve is for $\frac{4\kappa^2}{|\varepsilon_{L}|^2+|\varepsilon_{R}|^2}|\Phi_{-}|^2$.}
\end{center}
\end{figure}
Note that the normal modes for the matrix $A$ are related to $\delta \tilde{b}_{+}$ and $\delta \tilde{c}_{+}$ via $\Phi_{\pm}=(\delta \tilde{b}_{+}\pm\delta \tilde{c}_{+})/\sqrt{2}$. This is after we have removed the phase of $c_{s}$ via a redefinition of the variables. Then the relation (\ref{8}) leads to $\big(\frac{4\kappa^2}{|\varepsilon_{L}|^2+|\varepsilon_{R}|^2}\big)|\Phi_{\pm}|^2=\frac{4\kappa^2[(G\pm x)^2+4\kappa^2]}{|(2\kappa-ix)^2+G^2|^2}$. When $x=\sqrt{G^2-4\kappa^2}$, $\big(\frac{4\kappa^2}{|\varepsilon_{L}|^2+|\varepsilon_{R}|^2}\big)|\Phi_{\pm}|^2=\frac{1}{2}\big(1\pm\sqrt{1-\frac{4\kappa^2}{G^2}}\big)$. When $x=-\sqrt{G^2-4\kappa^2}$, $\big(\frac{4\kappa^2}{|\varepsilon_{L}|^2+|\varepsilon_{R}|^2}\big)|\Phi_{\pm}|^2=\frac{1}{2}\big(1\mp\sqrt{1-\frac{4\kappa^2}{G^2}}\big)$ . Therefore at the IEIT only one of the normal modes is excited. The dependence of the normalized modulus square of the normal modes $\frac{4\kappa^2}{|\varepsilon_{L}|^2+|\varepsilon_{R}|^2}|\Phi_{\pm}|^2$ for $G=6\kappa$ on the probe detuning $x$ is shown in Fig.~\ref{Fig5}. For $G=6\kappa$, the IEIT occurs at $x=\pm4\sqrt{2}\kappa$. From Fig.~\ref{Fig5}, we find that $\frac{4\kappa^2}{|\varepsilon_{L}|^2+|\varepsilon_{R}|^2}|\Phi_{+}|^2=0.971$ and $\frac{4\kappa^2}{|\varepsilon_{L}|^2+|\varepsilon_{R}|^2}|\Phi_{-}|^2=0.029$ at $x=4\sqrt{2}\kappa$,  $\frac{4\kappa^2}{|\varepsilon_{L}|^2+|\varepsilon_{R}|^2}|\Phi_{-}|^2=0.971$ and $\frac{4\kappa^2}{|\varepsilon_{L}|^2+|\varepsilon_{R}|^2}|\Phi_{+}|^2=0.029$ at $x=-4\sqrt{2}\kappa$. Thus only one of the normal modes is excited when the IEIT occurs. Finally we also mention that in the above analysis we have assumed that the pump detuning $\Delta$ is equal to the frequency $\omega_{m}$ of the mechanical oscillator. For $\Delta\neq\omega_{m}$, more general results including Fano profiles~\cite{Qu2013} would be obtained.

\section{Conclusions}

In conclusion, we have shown the IEIT in an optomechanical setup formed by a vibrating mirror within a Fabry-Perot cavity under the action of a strong coupling field. We show that the energy of the incoming probe fields is shared by the intracavity field and the movable mirror as they propagate through the optomechanical system. Further, we show that our optomechanical system can absorb the incident probe fields spanning a broad range of frequencies by varying the power and the frequency of the coupling field as long as the optomechanical interaction is not slower than the irreversible processes due to loss of photons out of the cavity mode $G\geq 2\kappa$. This property is clearly useful for filtering out frequencies for example from a probe pulse by using a pump with an appropriate spectral distribution. For the optomechanical system, the absorbed energy is stored in the coherent excitations of the mirror and the cavity, and thus the energy can be extracted by applying external perturbation fields or by suddenly decreasing the cavity quality factor $Q$. Finally we note that the study of the IEIT with quantized fields would be very interesting~\cite{Agarwalsq,Agarwalr,Eisaman,Appel,Lobino,Honda}. We hope to return to this question in a later paper.

\section*{Acknowledgments}
This work was supported by the Singapore National Research Foundation under NRF grant no.
NRF-NRFF 2011-07.\\

\noindent {\it Note added in proofs.} Although in the main body of the paper we assumed that there is no internal loss in the resonator. In practice many resonators suffer from the internal loss \cite{Wang1} say at the rate $\kappa_0$. Then the conditions for $\gamma_m$ and $x$ in Eq. (12) are modified by the replacement of $\kappa$ by $\kappa-\kappa_0/2$. The internal loss is then useful in the realization of the effect of this paper in cases where mirror damping is much smaller than $\kappa$.

\end{document}